\documentclass[%
 ,twocolumn%
 ,secnumarabic%
,amssymb, amsmath,nobibnotes, aps, prl]{revtex4}
\usepackage{epsfig}%
\usepackage{docs}%
\usepackage{bm}%
\expandafter\ifx\csname package@font\endcsname\relax\else
 \expandafter\expandafter
 \expandafter\usepackage
 \expandafter\expandafter
 \expandafter{\csname package@font\endcsname}%
\fi

\begin{document}

\def\G{GeV$^2$}
\def\Q{$Q^2$}
\def\v{\vspace{.1in}}
\def\F{\mathcal{F}}
\def\K{\mathcal{K}}

\title{\revtex~4 Prediction of Neutron Elastic Form Factors Using GPDs from Proton Elastic Form Factors and Isospin Symmetry}%

\author{Paul Stoler}%
\email{stoler@rpi.edu}
\affiliation {Physics Department, Rensselaer Polytechnic Institute, Troy, NY 12180}
\date{June 29, 2003}%

\begin{abstract}
The elastic neutron form factors $G_{En}$ and $G_{Mn}$ are calculated
in a GPD framework using GPDs obtained from fits to proton elastic
form factors $G_{Ep}$ and $G_{Mp}$, and isospin symmetry, with no
further changes in parameters.
The results  for $G_{En}$ are in good agreement with existing data, 
while those for $G_{Mn}$ are fair. The calculations predict
the form factors for future measurements at higher $Q^2$.
\end{abstract}



\maketitle

In recent years, the development of generalized parton distributions (GPDs)
~\cite{ji,rad_gpd,collins} has opened the  possibility of 
describing a great variety of
exclusive reaction in the multi GeV range in terms of a common nucleon 
structure. The constraints imposed by the description of many types of 
reactions offers the possibility of modeling the longitudinal and transverse
parton structure of nucleons.

Among the most direct consequences of the GPD formalism are the sum
rules which relate the various GPDs to the hadronic form
factors. Thus the proton elastic helicity conserving and helicity-flip
form factors may be written, respectively, as :  
 
$$F_{1p}(t)=\int^1_{-1}\sum_q  e_q H_p^q(x,\xi, t)dx$$  
$$F_{2p}(t)=\int^1_{-1}\sum_q  e_q E_p^q(x,\xi, t)dx$$

\noindent where $t = Q^2$ is the momentum transfer to the proton,
 $\xi$ is the longitudinal momentum transfer, and
$q$  signifies quark flavors. Without loss of generality  one
may work in a coordinate 
system in which the momentum transfer $t$ is transverse so that 
$\xi = 0$, and the GPDs may be written:

$$H_p^q(x,t) \equiv H_p^q(x,t,\xi=0)\ \ \ 
E_p^q(x,t) \equiv E_p^q(x,t,\xi=0)$$

Several authors~\cite{rad_wacs, kroll, kroll_GPD, burkardt}
have modeled the GPDs by Gaussian functions which
embody  general expected properties.
In particular,
$H_p^q(E,t=0) \to f_p^q(x)$, the unpolarized quark
distribution function and asymptotically  $H^q(E,-t\to\infty )$
narrows toward  $x=1$ (see \cite{burkardt,stoler_gpd}). In
terms of a Gaussian a simple model is,

\begin{equation}
H_p^q(x,t)=  f_p^q(x)e^{-\bar x t/4x \lambda_H^2}
\label{eq:Hsoft}
\end{equation}
 
\noindent in which $\bar x \equiv 1-x$. 
For $E_p^q(x,t)$ the we take the simple  ansatz.~\cite{afanasev}

\begin{equation}
E_p^q(x,t)=k_p^q(x)e^{-\bar x t/4x \lambda_E^2}.
\label{eq:Esoft}
\end{equation}
 
To account for  hard components of  $F_{1p}$  at $-t>10$  
ref.~\cite{stoler_gpd} modified the specific functional form for 
$H^q_p(x,t)$  and $E^q_p(x,t)$  as a Gaussian plus small power law shape 
in $-t$. \footnote{As in ref.~\cite{stoler_gpd}, eq.17, the parameter 
for the power law part of the GPD is $A_h=0.18$. Essentially no 
power law shape was required to fit  $F_{2p}$, where the data only goes up to 
$-t \sim 6$\ GeV$^2$/c$^2$.}

\begin{eqnarray} 
H_p^q(x,t)=f_p^q(x)exp(\bar x t /4x\lambda_H^2) + \cdot\cdot\cdot \\
E_p^q(x,t)=k_p^q(x)exp(\bar x t /4x\lambda_E^2) +  \cdot\cdot\cdot,
\label{eq:soft_hard}
\end{eqnarray}

\noindent in which  $\cdot\cdot\cdot$ indicates
the addition of small power law components in $-t$. 

To obtain  $E_p^u$ and
$E_p^d$,  needed for eqs.~\ref{eq:Hsoft} to \ref{eq:soft_hard} , 
the available data for $G_{Mp}$ and 
the recent JLab data~\cite{jones,gayou}  on $G_{Ep}/G_{Mp}$ were fit,
as in ref.~\cite{stoler_gpd}.

The  conditions at $t$=0 were also required, i.e.
$H_p(x,0)=e_uf_u(x)+e_df_d(x)$ and 
$E_p(x,0)=k_p^u(x)+k_p^d(x)$. 
The valence quark distribution functions  $f_p^u(x)$ and $f_p^d(x)$ 
are measured in DIS, and obtained from refs.~\cite{rad_wacs, martin}.
The functions $k_p^u(x)$ and $k_p^d(x)$ are not obtainable 
from evaluations of DIS . Following  ref.~\cite{stoler_delta}
the simple  phenomenological assumption
$k_p^q(x)\propto\sqrt{1-x}f_p^q(x)$ was used. This results in a satisfactory
ratio of  $F_{2p}/F_{1p}$, since for large $-t$, the quantity
$\sqrt{1-x}\to 1/\sqrt{-t}=1/Q$ with normalization obtained
by requiring the proton  $F_{2p}(0)=1.79$.

\begin{figure}[htb]
\centerline{\psfig{file=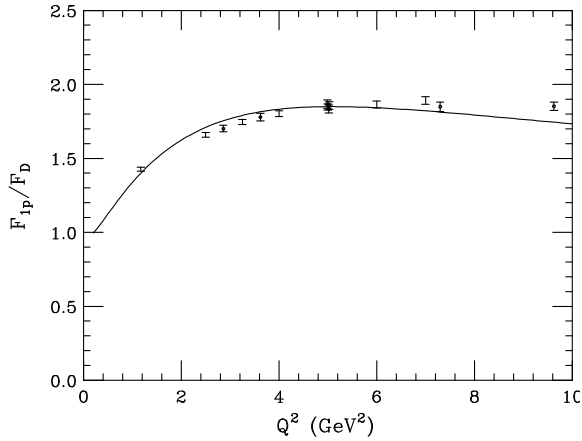,angle=90,width=3in}}
\caption{\label{fig1} Dirac form factor
$F_{1p}(Q^2)$ relative to the dipole  $G_D=1/(1+Q^2/.71)^2$.
The data are extracted  using the recent JLab data ~\cite{jones,gayou}
for $G_{Ep}/G_{Mp}$, and a recent reevaluation~\cite{brash}
of SLAC data of $G_{Mp}$~\cite{arnold}~\cite{andivahis}. 
The curve is the result  of the fit as discussed in the text.}
\end{figure}

\begin{figure}[htb]
\centerline{\psfig{file=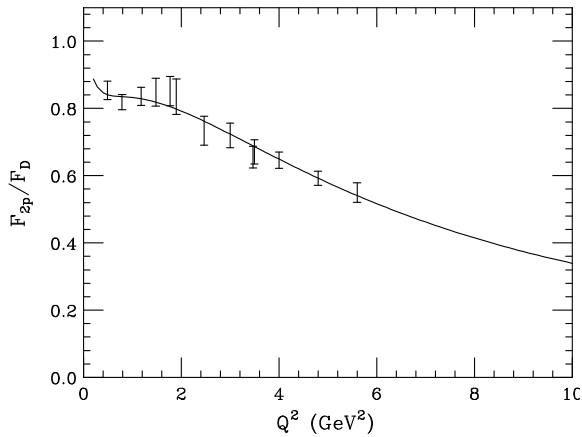,angle=90,width=3in}}
\caption{
\label{fig2} The Pauli form factor
$F_{2}/1.79F_D$ relative to the dipole  $F_D=1/(1+Q^2/.71)^2$.
The data are extracted using the recent JLab data~\cite{jones,gayou}
for $F_{2p}/F_{1p}$, multiplied by the fit curve for $F_{1p}/F_D$
shown in fig.~\ref{fig1}. The curve is the result 
of the simultaneous fit to the  $G_{Ep}/G_{Mp}$ and $G_{Mp}$
data as discussed in the text and fig.~\ref{fig1}  . }
\end{figure}

\noindent Adequate fits to the measured $G_{Mp}$ and $G_{Ep}/G_{Mp}$, or
equivalently  $F_{1p}$ and  $F_{2p}/F_{1p}$, were obtained
with $\lambda_H=0.76$ GeV/c and $\lambda_E=0.67$ GeV/c.
The results are shown in figs.~\ref{fig1} and \ref{fig2}.

This gives 

$$F_{1p}^u(0)= \int{ e_u H_p^u(x,0)}dx= \int{e_uf_p^u(x)dx} = 4/3$$
~$$F_{1p}^d(0)= \int{ e_u H_p^d(x,0)}dx= \int{e_df_p^d(x)dx} = -1/3$$

\noindent and

$$F_{2p}^u(0)\equiv \kappa_p^u = \int {e_u E_p^u(x,0)dx}=
\int{k_p^u(x)dx} = 1.67$$
$$F_{2p}^d(0)\equiv \kappa_p^d = \int {e_u E_p^d(x,0)dx}=
\int{k_p^d(x)dx} = -2.03$$ 

The neutron form factors were obtained from the fit to proton form factors by 
applying isospin symmetry.

       $$ H_n^u(x,t)=H_p^d(x,t)$$
       $$ H_n^d(x,t)=H_p^u(x,t)$$
       $$ E_n^u(x,t)=E_p^d(x,t)$$
       $$ E_n^d(x,t)=E_p^u(x,t)$$

\noindent and
 
 $$F_{1n}(t)=\int^1_{-1}\sum_q  e_q H_n^q(x, t)dx$$  
 $$F_{2n}(t)=\int^1_{-1}\sum_q  e_q E_n^q(x, t)dx$$

       $$ G_{En}=F_{1n}- \kappa_n\tau F_{2n} $$
       $$G_{Mn}=F_{1n}+\kappa_n F_{2n} $$

\noindent with $\kappa_n$ = -1.91 $\mu_N$.

\begin{figure}[h]
\begin{centering}
\centerline{\psfig{file=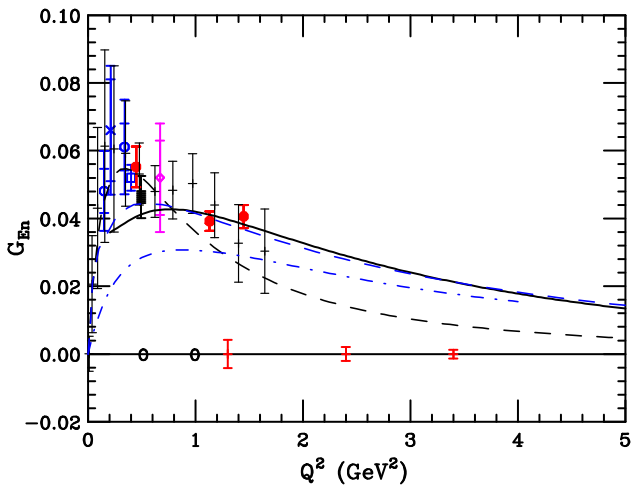, angle=90, height=2.5in}}
\caption{\label{fig3}The neutron electric form factor  $G_{En}$. 
 Data are plotted as follows. 
 Blue ($\times$)-ref.~\cite{Pa99},
open blue circles ($\circ$)-ref.~\cite{Os99,herberg}, filled blue 
triangle-ref.~\cite{Be99}, filled black square-ref.\cite{zhu}, magenta
diamond-ref.\cite{Ro99}. The black  pluses (+) are 
extracted from elastic $ed$ polarization and cross section measurements 
by ref.~\cite{Sc01}. Red filled circles-ref.~\cite{madey}.
The black open circles-ref~\cite{day} on the baseline represent
anticipated JLab Hall C data under analysis, and the red points along the 
baseline-ref.~\cite{Ca02} are projected data, including 
statistics for JLab approved experiment  in  JLab Hall A. 
The projections to higher  $Q^2$ planned for Hall A with
the 12 GeV upgrade will extend these measurements to at least  
$Q^2 \sim 5 $ GeV$^2$/c$^2$.
The solid black curve is the present prediction 
The blue dashed curve is due to ref.~\cite{miller} due to constituent
quarks and a pion cloud, while the blue dot-dash is from quarks only. 
The curve denoted by black dashes is the  Galster~\cite{galster} 
parameterization.}
\end{centering}
\end{figure}

The result for $G_{En}$ is shown in Fig.~\ref{fig3}.
The calculated form factor is somewhat lower than the existing data 
in the region $Q^2 =-t  < 0.75$ GeV$^2$/c$^2$, but accounts well for the
new JLab Hall C  data for   $Q^2 > 0.75$ \cite{zhu,madey}.
There is excellent agreement with the results of
a calculation of ref.~\cite{miller}, which is also shown in the figure.
The calculation of ref.~\cite{miller} uses a completely different framework, 
employing a relativistic constituent quark model with a pion cloud. 
For $G_{En}$ the pion cloud is important at small  $Q^2$, where the 
constituent quark contribution is very small.
However, for  $Q^2 > 1$\ GeV$^2$/c$^2$ the 
quarks become most important, with the role of the pion cloud diminishing.
 In the present calculation, the contribution of the sea quark pairs, which
presumably would mimic the pion cloud, was set to zero. 
The importance of a rigorously relativistic calculation of both the
constituent quarks and pion cloud is stressed in ref.~\cite{miller}.
For example, at high $Q^2$ the lower components of the Dirac
spinors, which introduce orbital angular momentum, become important.
The calculation of ref.~\cite{miller} employs several parameters,
however  the $Q^2$ dependence of the form factor at higher  $Q^2$ 
appears to be governed  more by relativistic effects than
the specific parameter set used. In particular a large number of sets of
these parameters can be found to give similar $Q^2$ dependence.

As seen in Fig.~\ref{fig3} both calculations give results at high $Q^2$ 
which lie above the Galster parameterization~\cite{galster}, as do the most
recent experimental data~\cite{madey}. This is not surprising since the 
Galster parameterization is simply an ad hoc fit to low  $Q^2$ data.

\begin{figure}[tbh]
\centerline{\psfig{file=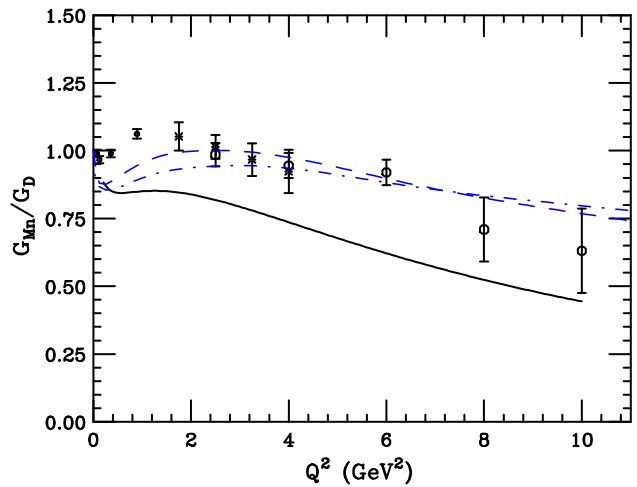, angle=90, height=2.5in}}
\caption{The neutron magnetic form factor $G_{Mn}$. The  data are as follows. 
Small circles for $Q^2 < 1$  GeV$^2$/c$^2$
ref.~\cite{kubon}. asterisks from  $Q^2 = $ 1.57 to 5  GeV$^2$/c$^2$
ref.~\cite{lung}. Large circles for  $Q^2 = $ 2.5 to 10  GeV$^2$/c$^2$
ref.~\cite{rock}.
The solid curve is the  prediction based on the present analysis.
The blue dashed and dot-dashed curves are due to ref.~\cite{miller}
for two different sets of parameters.}
\label{fig4}
\end{figure}

The result for $G_{Mn}$ are shown in Fig.~\ref{fig4}. Here, the fit to the 
experimental data is somewhat poorer than  for $G_{En}$. Also shown is the
result of the calculation of ref.~\cite{miller}. Curves are shown for
two of the many parameter sets which fit the data. A possible reason
for the better fit may be that ref.~\cite{miller} chooses parameters
in such a way that requires the fit to be rather good for all four
elastic form factors, while in the present case only the proton
form factors are fit, and then isospin symmetry is applied to obtain
the neutron form factors with the parameters fixed.

This note has pointed out the usefulness of GPDs in describing
elastic form factors. Alternatively, the elastic form factors, together
with isospin symmetry can be very important  for constraining
nucleon structure through GPDs. Further constraints  of details of nucleon
structure will be
possible by  including other high $-t$ experiments into the fit procedure.
These include high $W$ high $-t$ real and virtual Compton scattering,
and single meson photo and electroproduction, such as described in 
refs.~\cite{rad_wacs,kroll,kroll_huang}. It would be quite interesting 
if conceptual connections could be made between this technique and 
those of recent relativistic constituent quark models with a pion
cloud such 
as in ref.~\cite{miller}, or  recent helicity non-conserving pQCD based 
approaches~ref.~\cite{ralston, belitsky} which have had some success in 
explaining the $Q^2$ dependence of  $F_{2p}$.

{\bf Acknowledgments:}  
The work was partially supported by the {\em National Science Foundation}.


\begin{thebibliography}{16}

\bibitem{ji}X. Ji, {\em Phys. Rev. Lett.} {\bf 78}, 610 (1997). 

\bibitem{rad_gpd}A.V. Radyushkin, {\em Phys. Lett.} {\bf B380},417 (1996);
{\em Phys. Rev.} {\bf D56},5524 (1997).


\bibitem{collins}J. Collins, L. Frankfurt, and M. Strikman, {\em Phys. Rev.},
{\bf D56}, 2982 (1997).

\bibitem{rad_wacs}A.V. Radyushkin, {\em Phys. Rev.} {\bf D58},114008 (1998).


\bibitem{kroll} M. ~Diehl, Th. ~Feldmann, R. ~Jakob and P. ~Kroll,
{Eur. Phys.} {\bf C8}, 409 (1999); M. ~Diehl,
Th. ~Feldmann, R. ~Jakob and P. ~Kroll, {\em Nucl. Phys.} {\bf B596},
 33 (2001), Erratum-ibid. {\bf B605}, 647 (2001).

\bibitem{kroll_GPD}P. Kroll,  Proceedings of the Workshop on Exclusive Processes at 
High Momentum Transfer, A. Radyushkin and P. Stoler, eds., 
World Scientific, Singapore, 214 (2003), E-print: hep-ph/0207118.



\bibitem{burkardt} M. Burkardt,  Proceedings of the Workshop on Exclusive 
Processes at High Momentum Transfer, A. Radyushkin and P. Stoler, eds., 
World Scientific, Singapore, 99 (2003), and references within.


\bibitem{stoler_gpd}P. Stoler, {\em Phys. Rev.} {\bf D65}, 053013 (2002),hep-ph/0207312. 



\bibitem{afanasev}A. Afanasev, E-print: hep-ph/9910565;
 ``Proceeding of the JLAB-INT Workshop on Exclusive and
  Semi-Exclusive Processes at High Momentum Transfer'', 
  C. Carson and A. Radyushkin, eds. World Scientific (2000).
  May 1999

\bibitem{jones}M.K. Jones {\it et al.} {\em Phys. Rev. Lett.} {\bf 84},1398 (2000); 
\bibitem{gayou}O. Gayou {\it et al.} {\em Phys. Rev.} {\bf C64},038202 (2001).


\bibitem{martin}A.D. Martin et al., {\em Phys.Lett.}{\bf B53},216,(2002).

\bibitem{stoler_delta}P. Stoler, hep-ph/0210184.

\bibitem{brash}E.J. Brash  {\em et al.}, {\em Phys. Rev.} {\bf C65}, 051001(R) (2002).

\bibitem{arnold} R.G. Arnold  {\em et al.}, {\em Phys. Rev. Lett.} {\bf 57}, 174 (1986).

\bibitem{andivahis}L. Andivahis {\em et al.}, {\em Phys. Rev.} {\bf D50}, 5491 (1994).

\bibitem{zhu} H. Zhu {\it et al.}, Phys. Rev. Lett. {\bf 87}, 081801 (2001).

\bibitem{madey}R. Madey et al. {\it to be published}.


\bibitem{miller}G.~A.~Miller, Phys. Rev.{\bf c66}, 032201 (2002), nucl-th/0207007.

\bibitem{galster} S. Galster, H. Klein, J. Moritz, K.H. Schmidt, 
D. Wegener, Nucl. Phys. {\bf  B32}, 221 (1971).



\bibitem{Pa99} I. Passchier {\it et al.}, Phys. Rev. Lett. {\bf 82}, 4988 (1999). 

\bibitem{Os99} M. Ostrick {\it et al.}, Phys. Rev. Lett. {\bf 83}, 276 (1999). 
\bibitem{herberg} C. Herberg {\it et al.}, Eur. Phys. J. A {\bf 5}, 131 (1999).

\bibitem{Be99} J. Becker  {\it et al.}, Eur. Phys. J. A {\bf 6}, 329 (1999).


\bibitem{Ro99} D. Rohe  {\it et al.}, Phys. Rev. Lett. {\bf 83}, 4257 (1999).

\bibitem{Sc01} R. Schiavilla and I. Sick, Phys. Rev. C  {\bf 64}, 041002 (2001). 

\bibitem{day} Jefferson Lab Experiment E93-026, D. Day, G. Warren 
and M. Zeier, spokespersons, data under analysis. 

\bibitem{Ca02} Jefferson Lab Experiment E02-013, G. Cates, K. McCormick, 
B. Reitz  and B. Wojtsekhowski, spokepersons. 

\bibitem{kubon} G. Kubon et al.  Phys.Lett.B524:26-32,2002,nucl-ex/0107016.


\bibitem{lung}A. Lung {\it et al.},  Phys. Rev. Lett. {\bf70}, 718 (1993).

\bibitem{rock} S. Rock {\it et al.}, Phys. Rev. Lett. {\bf 49}, 1139 (1982). 


\bibitem{kroll_huang}H.W. Huang and P. Kroll,  Eur. Phys. J. {\bf C17}, w23 (2000).

\bibitem{ralston}Pankaj Jain and  John P. Ralston, hep-ph/0306194.

\bibitem{belitsky}A.V. Belitsky, X. Ji, F. Yuan hep-ph/0302043.


\end{thebibliography}
\end{document}